# Linear roto-antiferromagnetic effect in multiferroics: physical manifestations

Anna N. Morozovska[1], Victoria V. Khist[2], Maya D. Glinchuk[3], Venkatraman Gopalan[4*],

Eugene A. Eliseev [3†]

[1] Institute of Physics, National Academy of Sciences of Ukraine,

41, pr. Nauki, 03028 Kiev, Ukraine

[2] National University of Water Management and Nature Resources Use,

Rivne, 33028, st. Soborna, 11, Ukraine

[3] Institute for Problems of Materials Science, National Academy of Sciences of Ukraine,

3, Krjijanovskogo, 03142 Kiev, Ukraine

[4] Department of Materials Science and Engineering, Pennsylvania State University, University

Park, PA 16802, USA

**Abstract**

Using the theory of symmetry and the microscopic model we predicted the possibility of a linear roto-antiferromagnetic effect in the perovskites with structural antiferrodistortive and antiferromagnetic long-range ordering and found the necessary conditions of its occurrence. The main physical manifestations of this effect are the smearing of the antiferromagnetic transition and the jump of the specific heat near it. In the absence of external fields linear roto-antiferromagnetic coupling can induce a weak antiferromagnetic ordering above the Neel temperature, but below the temperature of antiferrodistortive transition. Therefore, there is the possibility of observing weak antiferromagnetism in multiferroics such as bismuth ferrite ($BiFeO_3$) at temperatures $T > T_N$, for which the Neel temperature $T_N$ is about 645 K, and the antiferrodistortive transition temperature is about 1200 K. By quantitative comparison with experiment we made estimations of the linear roto-antiferromagnetic effect in the solid solutions of multiferroic $Bi_{1-x}R_xFeO_3$ (R=La, Nd).

---

[*] Corresponding author: vgopalanpsu@gmail.com

[†] Corresponding author: eugene.a.eliseev@gmail.com

## 1. Introduction

Multiferroics, generally defined as ferroics with several types of long-range order interacting with each other, are unique model systems for fundamental physical studies of versatile couplings between the spontaneous polarization, magnetization, structural and antiferromagnetic order parameters [1, 2, 3, 4]. The most well-known and important effect for applications of multiferroics is the magnetoelectric coupling between the polarization and magnetization, through which one can write information by an electric field and readout it by a magnetic field [1, 4, 5].

Given the unique importance of multiferroics for wide variety of applications, other types of couplings are actively investigated in antiferrodistortive multiferroics in addition to the magnetoelectric coupling [6, 7, 8, 9, 10]. The couplings are associated with the presence of structural order parameter and its gradient. In the case of a inhomogeneous distribution of the order parameter, which is inevitable near the surface or in the presence of developed domain structure of ferroelectric, magnetic or structural types, there is a coupling between the various order parameters and their gradients [6-10]. Therefore, according to the theory of symmetry, the flexoelectric-antiferrodistortive and roto-flexoelectric coupling between structural, polar and magnetic order parameters [11, 12, 13] can exist in antiferrodistortive multiferroics in addition roto-magnetic and roto-electrical coupling [14, 15]. The prefix "roto" comes from the word rotation and indicates the static rotation of some atomic groups with respect to other parts of the crystal [14]. In the work the term "roto-symmetry" means only rotational symmetry of the oxygen octahedra $MO_6$ with respect to the cube $A_8$ in antiferrodistortive perovskites with the structural formula $AMO_3$. Oxygen atoms are displaced with respect to the centers of the cube faces $A_8$ in the antiferrodistortive phase, the angle or the value of the corresponding displacement is a structural order parameter (see e.g. [16]).

Coupling between the various orders parameters can be bilinear, linear-quadratic and biquadratic in the order parameters powers, depending on the extent to which the relevant parameter (or gradient) is proportional to the physical effect it has generated [1]. Biquadratic effects exist for arbitrary symmetry of multiferroic [17, 18, 19, 20], but the values of the corresponding tensor coupling constants strongly depend on its shape and size [21]. The appearance of non-zero bilinear effects is material-specific, they are determined by the spatial magnetic and roto- symmetry of the material (see e.g. [4, 7, 14]). Consequently bilinear effects are significantly less common, but their physical manifestations can be much more strong and non-trivial, rather than the manifestations of biquadratic effects [4]. Perhaps that is why researchers are actively "hunting" for bilinear coupling effects in multiferroics.

In this work we predicted the possibility of linear roto-antiferromagnetic coupling existence in perovskites with antiferrodistortive and antiferromagnetic ordering and found the necessary conditions of the coupling occurrence. Also we discuss the main physical manifestations of the effect, such as the smearing of antiferromagnetic (AFM) transition, specific heat jump near the transition and weak antiferromagnetism above the Neel temperature. We chose multiferroic bismuth ferrite (BiFeO$_3$) as the model material.

Our choice of BiFeO$_3$ is based on the fact that the material is one of the most promising multiferroic with a relatively high magnetoelectric coupling coefficient; it reveals antiferrodistortive order at temperatures below 1200 K; is ferroelectric with a high spontaneous polarization below 1100 K and antiferromagnetic below Neel temperature $T_N \approx (640 - 650)$ K [4, 22]. Despite the great amount of experimental studies on the BiFeO$_3$ multiferroic properties [23, 4, 5], many important issues remain unclear in the sense of understanding of the physical mechanisms responsible for the emergence and manifestation of these properties [22]. In other words, the theoretical description of BiFeO$_3$ physical properties is far behind the experiment. In particular *ab initio* calculations, which allow determining the parameters of antiferrodistortive and antiferromagnetic subsystems, magnetoelectric, roto-magnetic and roto-electric couplings of the corresponding long-range order parameters with each other in BiFeO$_3$, are absent to date. On the other hand reliable experimental results, which analyses, as we will show below, allow making conclusions about the exclusive importance of the roto-type couplings in BiFeO$_3$.

Before presenting the problem statement and original results, let us make some comments about chosen research methods. As we discuss the principal possibility of a new kind of interaction between two long-range order parameters (antiferrodistortive and antiferromagnetic) in the bulk of multiferroic, in order to establish the existence of a particular interaction between the order parameters and unambiguously define the structure of corresponding material tensor, one can use the theory of symmetry, if its spatial and magnetic symmetry group is known [14]. Functional form of the antiferrodistive-antiferromagnetic coupling contribution to the free energy and its effect on phase transitions in multiferroics can be established within the framework of the continuous medium mean-field Ginzburg-Landau theory [11, 13]. However, it is impossible to define the value of the coupling strength, i.e. to calculate non-zero coupling constant for a given material, using phenomenological approach and the theory of symmetry. One can estimate the strength of antiferrodistive-antiferromagnetic coupling either from first principle quantum mechanical calculations within a specific microscopic model, or from the fitting to experimental data. Both of these approaches are indispensable to determine the coupling constants and complement each other well, but by themselves they are not free from drawbacks. Most of the first principle calculations (such as carried out in the framework of DFT) do not take into account correctly inhomogeneous long-range depolarization electric field in ferroics and the totality of the structural and magnetic antiferrodistortive modes, as well as do not allow to say anything about the temperature dependence of the coupling constants. However it is possible to extract the coupling constant from the experiment sufficiently precisely and unambiguously, if the contribution of other effects is known within error margins.

In the work we consider step-by-step the microscopic picture necessary for the occurrence of linear roto-antiferromagnetic coupling in antiferrodistortive antiferromagnets with the structural formula $AMO_3$, establish the transformation law of linear roto-antiferromagnetic effect tensor, find its nonzero components of the theory of symmetry and estimate its numerical value from the smearing of specific heat jump near AFM transition for bismuth ferrite and its solid solutions $Bi_{1-x}R_xFeO_3$ (R=La, Nd).

**2. Microscopic model for bilinear roto-antiferromagnetic coupling (LRM) appearance**

The antiferromagnetic order parameter of the two-sublattice antiferromagnet is an axial vector **L**, that is equal to the difference of magnetization vectors of magnetic atoms in two equivalent sublattices A and B, $\mathbf{L} = (\mathbf{M}_A - \mathbf{M}_B)/2$, $\mathbf{M}_A = \sum_{i=1}^{A} g\mu_B \mathbf{S}_i$ и $\mathbf{M}_B = \sum_{i=1}^{B} g\mu_B \mathbf{S}_i$ (see **Figure 1a**). The antiferrodistortive order parameter is an axial vector, which is the angle of oxygen octahedra tilt $\varphi$. Below we use the equivalent form of order parameters defined as the oxygen displacement from symmetric position $\Phi = a \tan\varphi$, which can be calculated as the product of pseudocubic lattice constant $a$ with tangent of angle $\varphi$. As a rule, the angle $\varphi$ changes its sign in neighbouring cells, related to different sublattices A and B, namely $\mathbf{\Phi}_A = -\mathbf{\Phi}_B \equiv \mathbf{\Phi}$. The contribution of bilinear roto-magnetic coupling into the free energy is described be the following expression:

$$g_{RM}^{L} = \frac{\chi_{ij}}{2}\left(M_{Ai}\Phi_{Aj} + M_{Bi}\Phi_{Bj}\right) \equiv \frac{\chi_{ij}}{2}\left(M_{Ai} - M_{Bi}\right)\Phi_j \equiv \chi_{ij} L_i \Phi_j \qquad (1a)$$

Equation (1a) is invariant under the time inversion and the translation on the basic vector of pseudocubic lattice, since magnetization vectors of each sublattices $\mathbf{M}_j$ change their signs, and the sublattice A transforms into the sublattice B under such translation, therefore vector $\mathbf{\Phi}$ also changes its sign and from the macroscopic point of view nothing changes in the system. Thus the necessary condition for the linear roto-antiferromagnetic effect appearance is the simultaneous sign change of the vectors components $M_i$ and $\Phi_j$ in the neighbouring sublattices A and B. Otherwise the corresponding component $\chi_{ij}$ of the roto-antiferromagnetic tensor is identically zero in a high temperature parent phase, i.e. is becomes zero everywhere as it follows from the free energy expansion continuity on the irreducible representation of the parent phase. The same speculations leads us to the conclusion about impossibility of the nonzero linear roto-ferromagnetic term $\chi_{ij} M_i \Phi_j$ appearance.

Let us underline that the trilinear roto-antiferromagnetic coupling, described by invariants $\chi_{ijkl} L_i \Phi_j \Phi_k \Phi_l$ are $\tilde{\chi}_{ijkl} L_i L_j L_k \Phi_l$, should appear simultaneously with the bilinear roto-antiferromagnetic coupling considered above, as well as higher odd order couplings of the type. Below we will concentrate on the study of the bilinear roto-antiferromagnetic coupling (1) physical manifestations, since assume that the bilinear effect should dominate over the higher order odd ones under the same other conditions.

The transformation law of the linear roto-antiferromagnetic effect tensor components $\chi_{ij}$ under the point group symmetry operations with the matrix elements, $C_{ij}$, has the form $\chi_{ij} = (-1)^{tr} C_{im} C_{jk} \chi_{mk}$. The transformation laws of the order parameters are $\Phi_k = \det(C) C_{kf} \Phi_f$ и $L_i = (-1)^{tr} \det(C) C_{ip} L_p$. Determinant $\det(\mathbf{C}) = \pm 1$; the factor $tr$ denotes either the presence ($tr = 1$) or the absence ($tr = 0$) of the time-reversal operation coupled to the space transformation $C_{ij}$. Here the summation is performed over the repeating indexes.

For the magnetic and spatial symmetry groups corresponding to BiFeO$_3$ (spatial group is $R3\bar{c}$, magnetic group is $\bar{3}m$ or $I^-3_z^+2_x^-$ [24]), nonzero components of $\chi_{ij}$ are

$$\chi_{11}^{BiFeO3} = \chi_{22}^{BiFeO3} \neq \chi_{33}^{BiFeO3}. \tag{1b}$$

Nonzero components for EuTiO$_3$ are $\chi_{12}^{EuTiO3} = -\chi_{21}^{EuTiO3}$.

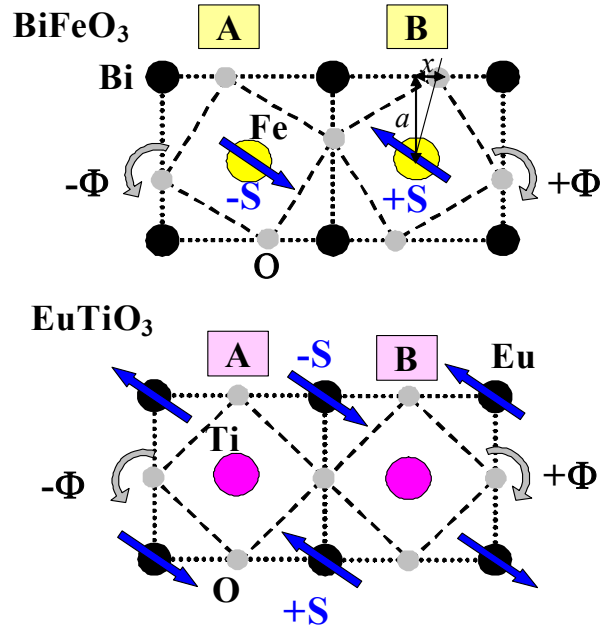

**Figure 1.** Schematic illustration to the microscopic model of the bilinear roto-antiferrodistortive coupling in the antiferrodistortive (AFD) phase of antiferromagnets with the structural formulae

AMO$_3$. The tilt **Φ** and spin **S** local values should be opposite for the neighbouring oxygen octahedrons, as shown in the figure for BiFeO$_3$ and EuTiO$_3$.

**3. Physical manifestations of the bilinear roto-antiferromagnetic coupling**

Description of an antiferrodistortive antiferromagnet in the framework of the phenomenological free energy approach shows, that the bilinear roto-antiferromagnetic coupling, in the manifestations we are interested in, does not influence the behaviour of magnetic and dielectric susceptibilities in external magnetic or electric fields, i.e. the coupling is unrelated with magnetoelectric effect. Thus we can consider an antiferrodistortive antiferromagnet in the absence of external fields for the purposes of the study. For the sake of simplicity we regard that both antiferrodistortive and antiferromagnetic phase transitions are of the second order.

The expression for the free energy density of the uniform antiferrodistortive-antiferromagnet in the absence of external magnetic and electric fields in the isotropic 1D-approximation has the following form

$$g = g_{AFM} + g_{AFD} + g_{RM}^L, \qquad (2a)$$

$$g_{AFM} = \frac{\alpha_L(T)}{2}L^2 + \beta_L \frac{L^4}{4}, \quad g_{AFD} = \frac{\alpha_\Phi(T)}{2}\Phi^2 + \frac{\beta_\Phi}{4}\Phi^4, \quad g_{RM}^L = \chi L \Phi \qquad (2b)$$

To describe the order parameters saturation behavior at low temperatures we used the quantum-corrected formula for the coefficient $\alpha_Q(T) = \alpha_{QT} T_Q \left( \coth(T_Q/T) - \coth(T_Q/T_q) \right)$ in Eqs.(2), which is valid in a wide temperature interval including low and high temperatures[25], where subscript $Q=L$ and the temperature $T_q$ is equal to Neel temperature $T_N$ of AFM parameter appearance; subscript $Q=\Phi$ and the temperature $T_q$ is equal to AFD transition temperature $T_S$ of the oxygen tilt appearance. At high temperatures $T \gg T_Q$ the formulae transforms into the classical limit, $\alpha_L(T) = \alpha_{LT}(T - T_N)$ and $\alpha_\Phi(T) = \alpha_{\Phi T}(T - T_S)$. Further let us solve the equations of state approximately in the assumption that the antiferrodistortive order parameter Φ appears at essentially higher temperatures $T_S$ than the temperature $T_N$ of spontaneous reversible antiferromagnetic order parameter $L$ appearance. This allows us to make a decoupling approximation on the coupling coefficient χ. This is by the way a typical situation realizing for e.g. Bi$_{1-x}$R$_x$FeO$_3$ (R=La, Gd, Nd, x=0 – 0.2) with $T_N \approx (635 - 655)$ K and $T_S \approx 1200$ K [4, 24], EuTiO$_3$ with $T_N \approx 5$ K and $T_S \approx 280$ K [26, 27, 28].

Equations of state, $\partial g/\partial L = 0$ and $\partial g/\partial \Phi = 0$, contains built-in fields $\chi\Phi$ and $\chi L$ correspondingly. As we have shown in the Supplement, the built-in fields leads to the antiferromagnetic (AFM) and antiferrodistortive (AFD) transition temperatures shifts, which are quadratic on the parameter χ, namely:

$$T_{AFM} = \frac{1}{2}(T_N + T_S) - \frac{1}{2}\sqrt{(T_S - T_N)^2 + \frac{\chi^2}{\alpha_{LT}\alpha_{\Phi T}}}, \quad (3a)$$

$$T_{AFD} = \frac{1}{2}(T_N + T_S) + \frac{1}{2}\sqrt{(T_S - T_N)^2 + \frac{\chi^2}{\alpha_{LT}\alpha_{\Phi T}}}. \quad (3b)$$

The shifts given by expressions (3) are relatively small under the validity of the strong inequality $\alpha_{LT}\alpha_{\Phi T}(T_S - T_N)^2 >> \chi^2$. Under the simultaneous validity of the later inequality and the condition $T_N << T_S$ typical for antiferrodistortive antiferromagnets, expressions (3) can be simplified to the form, $T_{AFM} \approx T_N - \chi^2/(4\alpha_{LT}\alpha_{\Phi T}T_S)$ and $T_{AFD} \approx T_S + \chi^2/(4\alpha_{LT}\alpha_{\Phi T}T_S)$.

Besides the shift (3), built-in field $\chi\Phi$ leads to the smearing of the AFM order parameter $L$ above the Neel transition temperature, i.e. in the paramagnetic phase. The smearing effect increases under the increase of $\chi$ value, as is shown in the **Figure 2a** by solid curves. Under the absence of linear coupling ($\chi=0$) one has $L = \pm\sqrt{\alpha_L(T)/\beta_L}$ at $T < T_N$. Structural order parameter is practically independent on $\chi$ and equal to $\Phi(T) \approx \pm\sqrt{\alpha_{\Phi T}(T)/\beta_\Phi}$ (see dotted curve in **Figure 2a**).

Unfortunately, the antiferromagnetic order parameter $L$ by itself is not directly observable, but some notion about its behavior can be obtained from the temperature dependences of neutron scattering and specific heat, if the contribution related with the long-range order appearance can be extracted. In particular the analyses and comparison with experiment of the specific heat changes $\delta C_P = -T\frac{d^2 g}{dT^2}$ allows us to verify the theoretical predictions made. In the typical case $T_{AFM} < T_{AFD}$ compact expression of the specific heat acquires the form:

$$\delta C_P = \begin{cases} -T\frac{d}{dT}\left(\frac{d\alpha_L}{dT}\frac{L^2}{2} + \frac{d\alpha_\Phi}{dT}\frac{\Phi^2}{2}\right), & T < T_{AFD}, \\ 0, & T > T_{AFD}. \end{cases} \quad (4)$$

As one can see from the **Figure 2b-c** the heat capacity variation peculiarity appeared in the vicinity of AFM transition, that is break at $\chi = 0$, which becomes smeared and shifted with $\chi$ increase. At $\chi = 0$ and temperatures $T < T_N$ the heat capacity change is associated with the appearance of AFM order parameter, $L \approx \pm\sqrt{\alpha_{LT}(T_N - T)/\beta_L}$, in the immediate vicinity below the AFM order phase transition. At $\chi = 0$ and temperatures $T_N < T < T_S$ the parameter $L = 0$. Therefore under the absence of bilinear coupling between the sublattices magnetization and antiferrodistortive tilts, only the sharp jump appears on the specific heat at Neel temperature $T_N$. The jump value is equal to $\delta C_P^N = T_N (\alpha_{LT})^2 / 2\beta_L$.

Note that linear magnetoelectric effect (if one exists in a concrete antiferromagnet) does not contribute into the specific heat behaviour in the absence of external fields, and so the question about the contribution of other coupling, biquadratic e.g. roto-electric, magnetoelectric or roto-magnetic, to the specific heat smearing near $T_N$ arises. If these or others contributions exist how they can be separated from the ones caused by the considered bilinear coupling? In order to answer the question let us estimate the contribution of the biquadratic couplings between different order parameters and their mean squire fluctuations into the specific heat of antiferrodistortive ferroelectric-antiferromagnet.

In order to calculate the contribution one can modify the free energy (2) by adding the ferroelectric contribution, $g_{FE} = \alpha_P(T)P^2 + \beta_P P^4$, and biquadratic couplings, $g_{BQ} = \xi\Phi^2 L^2 + \eta\Phi^2 P^2 + \lambda P^2 L^2$. Ferroelectric polarization leads to the occurrence of additional term in Eq.(4), that is equal to $\delta C_P \approx -T \dfrac{d}{dT}\left(\dfrac{d\alpha_P}{dT}\dfrac{P^2}{2}\right)$ and nonzero in ferroelectric phase at temperatures $T < T_{FE}$.

Rigorously speaking the biquadratic terms can only shift corresponding transition temperatures, but cannot lead to any smearing of diffuseness in the transition region. Thus the smearing effect is related with the bilinear term $\chi L \Phi$ in thermodynamic limit.

Using Ginzburg-Levanuk approach [29] for the estimation of the order parameters mean square fluctuations contribution into the specific heat in the vicinity of AFM phase transition, we include the gradient terms in the free energy (2), which have the simplest form in the isotropic approximation, $g_{gr} = \dfrac{\gamma_\Phi}{2}(\nabla\Phi)^2 + \dfrac{\gamma_P}{2}(\nabla P)^2 + \dfrac{\gamma_L}{2}(\nabla L)^2$, and the entropy that density near AFM phase transition is approximately equal to $g_{fl}^{AFM} \approx \dfrac{k_B T}{2\pi^2} \int_0^{k_{max}} k^2 \ln(\alpha_{LT}|T - T_{AFM}| + \gamma_L k^2) dk$. Corresponding expression for the order parameters fluctuations contribution into the specific heat change near AFM has the form [29]:

$$\delta C_P^{fl} \approx \dfrac{k_B T^2 \alpha_{LT}^{3/2}}{8\pi \gamma_L^{3/2} \sqrt{|T - T_{AFM}|}} \quad (5)$$

Expression (5) diverges at $T = T_{AFM}$ for finite $\gamma_L$ due to the fluctuations, which contribution disappears in the limiting case $\gamma_L \to \infty$ (thermodynamic limit of Landau theory). Elementary estimations made for the typical values of parameters $\alpha_{LT}$, $\beta_L$ и $\gamma_L$ [30], have shown, that the smearing effect defined in Eq. (5), is essential only in very narrow vicinity of antiferromagnetic phase transition (for temperatures interval from fractions of Kelvin to few Kelvin wide), while the experimentally observed range of specific heat jump smearing is of order of 20-50 Kelvin. Really, at χ=0 the ratio $\frac{\delta C_P^{fl}}{\delta C_P^N} = \frac{k_B T^2 \beta_L}{4\pi \gamma_L^{3/2} T_N \sqrt{\alpha_{LT} |T - T_N|}}$ becomes less than 0.01 already at $|T - T_N| > 1$ K. Consideration of χ-effect Barret's law for temperature dependence of $\alpha_L(T)$ could not change significantly this estimation. Therefore, fluctuation (5) does not make a significant contribution to the smearing of the specific heat jump, observable in experiment, approving the conclusion that most of the smearing of AFM phase transition is associated with bilinear coupling χLΦ only.

Below we consider multiferroic $BiFeO_3$, which is antiferrodistive ferroelectric – antiferromagnet with critical temperatures $T_N \approx 645$ K, $T_C \approx 1100$ K and $T_S \approx 1200$ K [22]. Our fitting of temperature dependence of AFM order parameter L in $BiFeO_3$ obtained from neutron scattering experiment of Fischer et al [31] is shown **Figure 2a**. The fitting of temperature dependence of specific heat and its part associated with transition to AFM phase is shown in **Figure 2b-c** for the experimental results of Kallaev et al [32]. Solid curves from Fig.2 correspond to nonzero value of effective parameter $\tilde{\chi} = (\chi/\beta_L)\sqrt{\alpha_{\Phi T}/\beta_\Phi}$ =2, 5, 10 SI units, dashed curves correspond to the case $\tilde{\chi} = 0$. The best fitting was obtained for $\tilde{\chi} = 2$ SI units. It is clearly seen that abrupt jumps of order parameter L temperature dependences, corresponding to the calculation at $\tilde{\chi} = 0$, is in a satisfactory agreement with experimental points below $T_N$. The same situation is for the curve calculated at $\tilde{\chi} = 2$ SI units, for which the small L exists above the Neel temperature, decreases with temperature increases and tends to zero at $T \to T_S$. However the specific heat features observed at $T = T_N$ are evidently blurred in the temperature region $T_N < T < 1.1 T_N$, at that the smearing can be satisfactory described only at $\tilde{\chi} = 2$ SI units, but not at $\tilde{\chi} = 0$. Therefore we can conclude that the diffuse "tail" of the antiferromagnetic order parameter L (as shown in the **Figure 2a**), that exists at nonzero $\tilde{\chi}$ values, pointed out on the possibility of the "weak" improper antiferromagnetism induced by the antiferrodistortive structural ordering.

However nothing definite can be concluded about the $\tilde{\chi}$ value from the analyses of the experimental data shown in the **Figure 2a**, because it is extremely complex to register relatively small sublattices magnetization values by neutron scattering. On the other hand it is evident from the specific heat behavior shown in the **Figure 2b-c**, that the inequality $|\tilde{\chi}| > 1$ SI units is necessary for the satisfactory agreement with experiment.

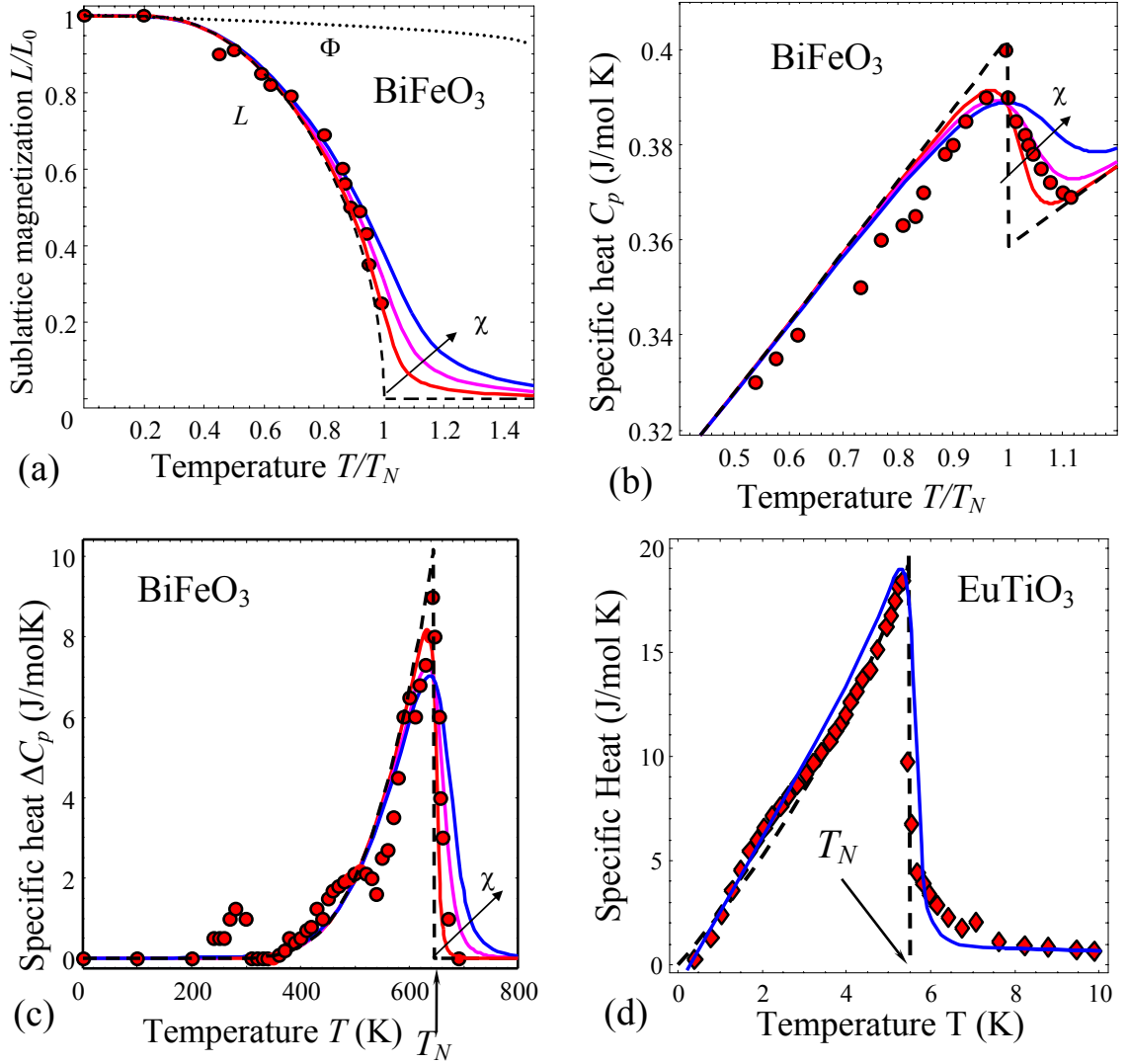

**Figure 2. (a)** Sublattices magnetization $L(T)/L(0)$ and **(b)** heat capacity variation as a function of reduced temperature $T/T_N$. Symbols are experimental data for BiFeO$_3$ from Fischer et al [31] on neutron scattering and Kallaev et al [32] for specific heat correspondingly. Curves are calculated by us for effective coupling constant $\tilde{\chi}$ = 2, 5, 10 SI units (solid curves) and $\chi$ = 0 (dashed curve); $T_N$ = 645 K, $T_L$ = 550 K, $T_\Phi$ = 100 K, $T_S$ = 1200 K. Dotted curve in the plot **(b)** is the AFD order parameter, $\Phi/\Phi_0$, that is almost independent on $\chi$ value for chosen parameters. **(c)** Temperature dependencies of the anomalous contribution to the BiFeO$_3$ specific heat. Symbols are experimental data from [32]. Solid curves are calculated at $\tilde{\chi}$ = 2, 5, 10 SI units, dashed curves corresponds to $\chi$ = 0. **(d)** Specific heat variation of EuTiO$_3$ near the AFM transition. Symbols represent the experimental data [26-28]. Effective coupling constant $\chi$ = 0 (dashed curve) and $\tilde{\chi}$ = 2 SI units (solid curve); $T_N$ = 5.5 K, $T_S$ = 285 K. Other parameters of EuTiO$_3$ are listed in the Table 1 in the ref [15].

Heat capacity variation of EuTiO$_3$ near the AFM transition is shown for comparison in the **Figure 2d**. As one can see from the plot nonzero $\tilde{\chi}$ (solid curve) describes the experimental data better than $\chi = 0$ (dashed curve). It is worth to underline that the smearing of sublattice magnetization and specific heat for BiFeO$_3$ and EuTiO$_3$ shown in **Figs.2** looks like the smearing of ferroelectric properties in external electric field [33]. This obviously confirmed the statement that the terms $\chi\Phi$ and $\chi L$ can be considered as built-in fields in the lattices.

In the next section we show how the existence of roto-antiferromagnetic coupling can be approved from the specific heat behavior in the BiFeO$_3$-based solid solutions, and estimate the effect value.

## 4. Determination of the roto-antiferromagnetic coupling constant for Bi$_{1-x}$R$_x$FeO$_3$ solid solutions

Available experimental results demonstrate noticeable features of the temperature dependencies of the specific heat in Bi$_{1-x}$R$_x$FeO$_3$ (R=La, Nd, x=0 – 0.2) solid solutions [34]. The features appears at the temperature of the antiferromagnetic phase transition that is about (640-650) K. Corresponding experimental results are shown by symbols in the **Figures 3.** As one can see from the figure dashed curves calculated at $\tilde{\chi} = 0$ and different composition x do not describe the specific hear smearing at temperatures $T > T_N$. Solid curves, calculated as $\tilde{\chi} = (2 - 2.5)$ SI units and $T_N = (645 - 655)$ K in dependence of x, describe the smearing effect adequately, proving the importance of the bilinear roto-antiferromagnetic effect for the understanding of the specific heat behaviour near the antiferromagnetic phase transition. The inclusion of the bilinear roto-antiferromagnetic effect is necessary for the quantitative description of the experimental data.

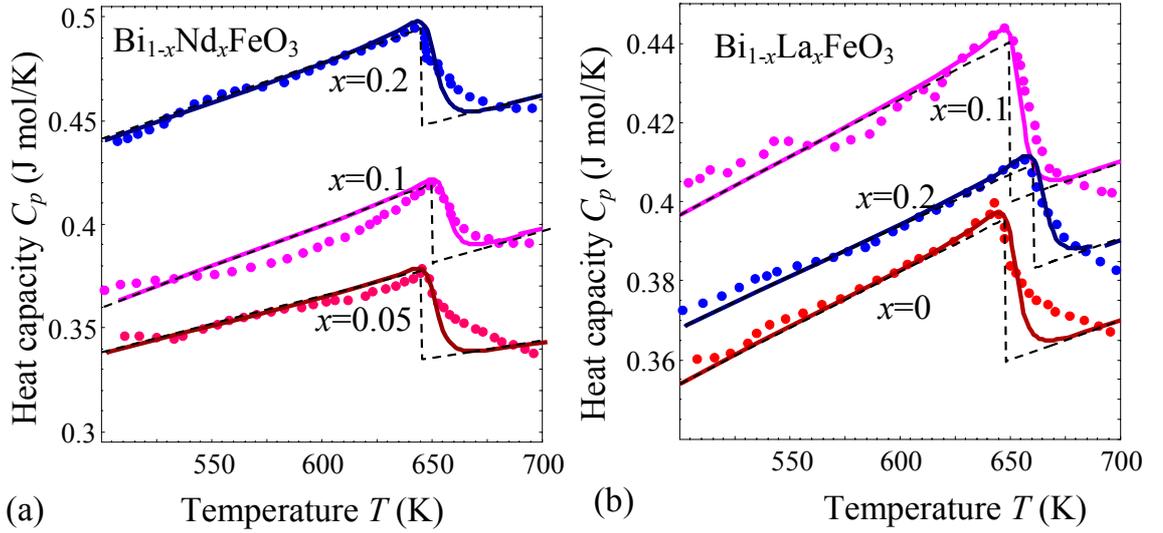

**Figure 3.** Temperature dependence of the specific heat near AFD phase transition of the solid solutions $Bi_{1-x}R_xFeO_3$ (R=La, Nd, x= 0 – 0.2). Symbols are experimental data for $Bi_{1-x}R_xFeO_3$ from Amirov et al [34] for heat capacity correspondingly. Dashed curves are calculated by us for dimensionless coupling constant $\tilde{\chi} = 0$. Solid curves correspond to different nonzero $\tilde{\chi} = (2 - 2.5)$ SI units and $T_N = (645 - 655)$ K depending on the composition $x$, $T_L = 550$ K, $T_\Phi = 100$ K, $T_S = 1200$ K.

Notice that we did not aimed to determine all material parameters of $Bi_{1-x}R_xFeO_3$ from the fitting of the specific heat variation $\delta C_p(T)$ and normalized antiferromagnetic order parameter $L(T)/L(0)$ temperature dependences, because it was impossible. Only the ratio $(\alpha_{LT})^2/2\beta_L$ can be defined from specific heat jump $\delta C_p^N$, and the temperature dependence of the ratio $\alpha_L(T)/\alpha_{LT}$ can be determined from the temperature dependence $L(T)/L(0)$. In order to define the value of $\alpha_{LT}$ one needs the temperature dependence of the antiferromagnetic susceptibility that we could not found in literature. Despite the difficulty we reached our goal and found the effective value of the roto-antiferrodistortive coupling constant, $\tilde{\chi} = (\chi/\beta_L)\sqrt{\alpha_{\Phi T}/\beta_\Phi}$, from the fitting of experimental data.

## 5. Conclusions

The possibility of the linear roto-antiferromagnetic effect existence in perovskite-multiferroics with the structural formula $AMO_3$, antiferrodistive and antiferromagnetic ordering is demonstrated. Within the framework of the theory of symmetry and the microscopic model the necessary conditions for this effect occurrence are the simultaneous change in the sign of the corresponding components of the elementary magnetization vectors in the neighboring antiferromagnetic sublattices coupled with the change of the antiferrodistive displacement direction in the neighboring oxygen octahedron $MO_6$. Let us underline that the trilinear roto-antiferromagnetic coupling should appear simultaneously with the considered bilinear roto-antiferromagnetic coupling, as well as higher odd order couplings of the type.

Physical manifestations of roto-antiferromagnetic effect is smearing of the antiferromagnetic phase transition and the emergence of small "improper" antiferromagnetic order parameter $L$ above the Neel temperature and below the temperature of antiferrodistortive transition. The parameter $L$ is induced by the product of the AFD order parameter $\Phi$ on the roto-antiferromagnetic coupling constant $\chi$, at that the term $\chi\Phi$ acts as effective built-in conjugated field for the parameter. Therefore, there is the possibility to observe weak "improper" antiferromagnetism induced by the structural antiferrodistive ordering $\Phi$ above the Neel temperature. For example, in bismuth ferrite, for which the antiferromagnetic transition temperature is of the order of (640-655) K, and the temperature antiferrodistortive transition goes above 1200 K, the temperature dependence of $L$ was measured by neutron scattering method. However, the available experimental data cannot say anything definite about the value of a smearing effect, because it is extremely difficult to register a sufficiently small value of $L$ by neutron scattering.

Roto-antiferromagnetic effect also leads to the smearing of the jump of the specific heat near the temperature of antiferromagnetic phase transition. By quantitative comparison with experiments we made estimates of the roto-linear effect in antiferromagnetic solid solutions of multiferroic $Bi_{1-x}R_xFeO_3$ and determine the optimal value of the roto-antiferromagnetic coupling constants from the fitting to experimental data. Calculated dependencies describe the experiment quite satisfactory, thus proving the importance of bilinear roto-antiferromagnetic effect for the understanding of the mechanisms responsible for the temperature dependence of the specific heat behaviour near the antiferromagnetic phase transition.

# References

1 Manfred. Fiebig, "Revival of the magnetoelectric effect." *Journal of Physics D: Applied Physics* 38, no. 8: R123 (2005).

2 N. A. Spaldin and M. Fiebig, Science 309, 391-392 (2005).

3 J.M. Rondinelli, N.A. Spaldin. Advanced Materials.23, 3363–3381 (2011).

4 A. P., Pyatakov, A. K.Zvezdin, Magnetoelectric and multiferroic media. *Physics-Uspekhi*, *55*(6), 557-581 (2012).

5 J. F. Scott, "Data storage: Multiferroic memories." *Nature materials* 6, no. 4: 256-257 (2007).

6 V.G. Bar'yakhtar, V.A. L'vov, D.A. Yablonskii. JETP Lett. 37, 673 (1983).

7 Maxim Mostovoy, Phys. Rev. Lett. 96, 067601 (2006).

8 A. Sparavigna, A. Strigazzi, and A. Zvezdin, Phys. Rev. B 50, 2953 (1994).

9 A.P. Pyatakov and A.K. Zvezdin, Eur. Phys. J. B 71, 419–427 (2009).

10 B.M. Tanygin. Journal of Magnetism and Magnetic Materials 323, 616–619 (2011).

11. E.A. Eliseev, M.D. Glinchuk, V. Khist, V.V. Skorokhod, R. Blinc, A.N. Morozovska. Linear magnetoelectric coupling and ferroelectricity induced by the flexomagnetic effect in ferroics. Phys.Rev. B 84, 174112 (2011)

12. Morozovska, Anna N., Eugene A. Eliseev, Maya D. Glinchuk, Long-Qing Chen, and Venkatraman Gopalan. "Interfacial polarization and pyroelectricity in antiferrodistortive structures induced by a flexoelectric effect and rotostriction." Physical Review B 85, no. 9 (2012): 094107.

13. Eugene A. Eliseev, Sergei V. Kalinin, Yijia Gu, Maya D. Glinchuk, Victoria Khist, Albina Borisevich, Venkatraman Gopalan, Long-Qing Chen, and Anna N. Morozovska. Universal emergence of spatially-modulated structures induced by flexo-antiferrodistortive coupling in multiferroics. Phys.Rev. B 88, 224105 (2013).

14 Venkatraman Gopalan, and Daniel B. Litvin. "Rotation-reversal symmetries in crystals and handed structures." *Nature materials* 10, no. 5: 376-381(2011).

15 Eugene A. Eliseev, Maya D. Glinchuk, Venkatraman Gopalan, Anna N. Morozovska. Rotomagnetic couplings influence on the magnetic properties of antiferrodistortive antiferromagnets http://arxiv.org/abs/1409.7108

16. Hiromoto Uwe, and Tunetaro Sakudo. *Physical Review B* 13, no. 1: 271 (1976).

17. M. J. Haun, E. Furman, T. R. Halemane and L. E. Cross, Ferroelectrics **99**, 55 (1989)

18. B. Houchmanzadeh, J. Lajzerowicz and E Salje, J. Phys.: Condens. Matter **3**, 5163 (1991)

19. E.V. Balashova and A.K. Tagantsev, Phys. Rev. B 48, 9979 (1993).

20. A.K. Tagantsev, E. Courtens and L. Arzel, Phys. Rev. B, **64**, 224107 (2001).

21. M.D. Glinchuk, E.A. Eliseev, A.N. Morozovska, R. Blinc. Phys. Rev. B 77, № 2, 024106-1-11 (2008)

22 Gustau Catalan and James F. Scott, Adv. Mater. **21**, 1–23 (2009)

23 G. A. Smolenskii, L. I. Chupis, Sov. Phys. Usp. **25**, 475 (1982).

24 Кадомцева, Звездин, Письма в ЖЭТФ, **79**, 705 (2004)

25 J. H. Barrett, Phys. Rev. **86**, 118 (1952).

26 T. Katsufuji and H. Takagi, Phys. Rev. B **64**, 054415 (2001).

27 V. V. Shvartsman, P. Borisov, W. Kleemann, S. Kamba, T. Katsufuji. Phys. Rev. B **81**, 064426 (2010).

28 P. G. Reuvekamp, R. K. Kremer, J. Kohler, and A. Bussmann-Holder. Phys. Rev. B **90**, 094420 (2014)

29 B. A. Strukov and A. P. Levanyuk, Ferroelectric Phenomena in Crystals (Springer, Berlin, 1998)

30 {$k_B$=1.38×10$^{-23}$ J/K, $\alpha_{TP}$ = 4.366·10$^{-7}$, $T_N$ = 650 K, $\beta_P$ = 0.323*10$^{-17}$, $\gamma_{LP}$ = 10$^{-8}$ m$^3$/F}

31 P. Fischer, M. PoIomska, I. Sosnowska, M. Szymanski J. Phys. C: Solid St. Phys., 13, 1931-40(1980).

32 С.Н. Каллаев et al, Теплоемкость и диэлектрические свойства мультиферроиков $Bi_{1-x}Gd_xFeO_3$ (x = 0−0.20). Физика твердого тела, том 56, вып. 7, 1360-1363 (2014)

33 Lines, Glass, textbook

34 Амиров et al, $Bi_{1-x}Re_xFeO_3$ (Re=La, Nd; x=0-0,2) (2010)

**Supplementary Information**

The approximate expression for the free energy of the antiferrodistortive-antiferromagnet in the absence of external magnetic (H=0) and electric (E=0) fields has the following form

$$g = g_{AFM} + g_{Coupling} + g_{AFD}, \qquad (S.1a)$$

$$g_{AFM} = \frac{\alpha_L(T)}{2}L^2 + \beta_L \frac{L^4}{4}, \quad g_{AFD} = \frac{\alpha_\Phi(T)}{2}\Phi^2 + \frac{\beta_\Phi}{4}\Phi^4 \qquad (S.1b)$$

$$g_{Coupling} = \chi_{ij} L_i \Phi_j + \frac{1}{2}\xi_{RM}^L L^2 \Phi^2, \qquad (S.1c)$$

Temperature-dependent coefficients are $\alpha_L(T) = \alpha_{LT} T_L (\coth(T_L/T) - \coth(T_L/T_N))$ and $\alpha_\Phi(T) = \alpha_{\Phi T} T_\Phi (\coth(T_\Phi/T) - \coth(T_\Phi/T_S))$. Further let us solve the equations of state approximately in the assumption that the antiferrodistortive order parameter $\Phi$ appears at essentially higher temperatures $T_S$ than the temperature $T_N$ of spontaneous reversible antiferromagnetic order parameter $L$ appearance. This allows us to make a decoupling approximation on the coupling constants $\xi_{RM}^L$ and $\chi$. This is by the way a typical situation realizing for e.g. BiFeO$_3$, EuTiO$_3$, etc. In a scalar approximation equations of state for the order parameter determination are

$$\frac{\partial g}{\partial L} = \alpha_L L + \beta_L L^3 + \xi_{RM}^L L \Phi^2 + \chi \Phi = 0, \qquad (S.2a)$$

$$\frac{\partial g}{\partial \Phi} = \alpha_\Phi \Phi + \beta_\Phi \Phi^3 + \xi_{RM}^L L^2 \Phi + \chi L = 0. \qquad (S.2c)$$

The phase transition point of the long-range order appearance can be found from the condition of linearized system $\alpha_{LT}(T - T_N)L + \chi\Phi = 0$, $\alpha_{\Phi T}(T - T_S)\Phi + \chi L = 0$ nonzero determinant, which is $\alpha_{LT}(T - T_N)\alpha_{\Phi T}(T - T_S) - \chi^2 = 0$. Corresponding rigorous and approximate expressions and their for transition temperatures become

$$T_{AFM} = \frac{1}{2}(T_N + T_S) - \frac{1}{2}\sqrt{(T_S - T_N)^2 + \frac{\chi^2}{\alpha_{LT}\alpha_{\Phi T}}} \approx T_N - \frac{\chi^2}{4\alpha_{LT}\alpha_{\Phi T}T_S} \qquad (S.3a)$$

$$T_{AFD} = \frac{1}{2}(T_N + T_S) + \frac{1}{2}\sqrt{(T_S - T_N)^2 + \frac{\chi^2}{\alpha_{LT}\alpha_{\Phi T}}} \approx T_S + \frac{\chi^2}{4\alpha_{LT}\alpha_{\Phi T}T_S} \qquad (S.3b)$$

One can see from Eqs.(S.3) hat the correction to transition temperatures is proportional to the squire of the coupling constant $\chi$. In the decoupling approximation

$$\Phi_S(T) \approx \pm\sqrt{-\frac{\alpha_\Phi(T)}{\beta_\Phi}} \qquad (S.4)$$

and the term $\chi L$ can be neglected in Eq.(S.2b). Then Eq.(S.2a) immediately transforms into the equation for L in the some built-in field $\chi \Phi_S$, namely:

$$\left(\alpha_L + \xi_{RM}^L \Phi_S^2\right)L + \beta_L L^3 = -\chi \Phi_S. \tag{S.5a}$$

The solution of the equation that has physical sense at positive $\chi \Phi_S > 0$, has the following form

$$L(T) = \frac{\sqrt[3]{2}(\Delta(\Phi_S,\chi))^2 - 2\sqrt[3]{3}\left(\alpha_L + \xi_{RM}^L \Phi_S^2\right)\beta_L}{\sqrt[3]{36}\Delta(\Phi_S,\chi)}. \tag{S.5b}$$

Here the function $\Delta(\Phi_S,\chi) = \left(9\beta_L^2 \chi \Phi_S + \sqrt{3\beta_L^3 \left(4\left(\alpha_L + \xi_{RM}^L \Phi_S^2\right)^3 + 27\beta_L (\chi \Phi_S)^2\right)}\right)^{1/3}$. Expression (S.5b) means, that the AFM phase transition becomes diffuse (see **Figure 2a**). Heat capacity variation related with AFD-AFM modes can be calculated as the second derivative of the free energy on temperature $T$, multiplied by $T$, namely in our approximations:

$$\begin{aligned}
\delta C_V &= -T\frac{d^2 g}{dT^2} \equiv -T\frac{d}{dT}\left(\frac{\partial g}{\partial \Phi}\frac{d\Phi}{dT} + \frac{\partial g}{\partial L}\frac{dL}{dT} + \frac{\partial g}{\partial T}\right) \\
&\equiv -T\frac{d}{dT}\left(\frac{\partial g}{\partial T}\right) \equiv -T\frac{d}{dT}\left(\frac{d\alpha_L}{dT}\frac{L^2}{2} + \frac{d\alpha_\Phi}{dT}\frac{\Phi^2}{2}\right) \\
&\equiv \begin{cases} -T\left(\frac{d\alpha_L}{dT}\frac{dL}{dT}L + \frac{d^2\alpha_L}{dT^2}\frac{L^2}{2} + \frac{d\alpha_\Phi}{dT}\frac{d\Phi}{dT}\Phi + \frac{d^2\alpha_\Phi}{dT^2}\frac{\Phi^2}{2}\right), & T < T_{AFD}, \\ 0, & T > T_{AFD}. \end{cases}
\end{aligned} \tag{S.6}$$